\documentclass[epj,nopacs]{svjour}

\usepackage[T1]{fontenc} 

\newcommand{\bran}{\mbox{$\cal B$}}
\newcommand{\gchfb}     {\mbox{$R_c \cdot f(c\rightarrow H_c) \cdot \bran$}}
\newcommand{\gbhfb}     {\mbox{$R_b \cdot f(b\rightarrow H_c) \cdot \bran$}}

\newcommand{\dsp}        {\mbox{$D^{\ast +}$}}

\newcommand{\dssp}       {\mbox{$D_s^+$}}

\newcommand{\dz}         {\mbox{$D^{0}$}}
\newcommand{\dc}         {\mbox{$D^+$}}

\newcommand{\lcp}        {\mbox{$\Lambda_c^+$}}

\newcommand{\fcds}       {\mbox{$f(c \rightarrow D^{\ast +})$}}
\newcommand{\fbds}       {\mbox{$f(b \rightarrow D^{\ast +})$}}
\def\dsdz{ {\dsp}~\rightarrow~\dz~\pi^{+}}
\def\dssphipi{ {\dssp}~\rightarrow~\phi~\pi^{+}%
        \rightarrow~(K^{-}~K^{+})~\pi^{+} }
\def\dzkpi{ {\dz}~\rightarrow~K^{-}~\pi^{+} }
\def\dckpipi{ {\dc}~\rightarrow~K^{-}~\pi^{+}~\pi^{+} }
\def\lcpkpi{ {\lcp}~\rightarrow~p~K^{-}~\pi^{+} }

\title{\boldmath Fragmentation fractions of $c$ and $b$ quarks into charmed hadrons at LEP}

\author{L. Gladilin}

\institute{D.V.Skobeltsyn Institute of Nuclear Physics, M.V.Lomonosov Moscow State University,\\1(2), Leninskie gory, GSP-1, Moscow 119991, Russian Federation,
\email{gladilin@mail.cern.ch}}

\abstract{
The fragmentation fractions of  
$c$ and $b$ quarks into
the weakly decaying
charmed hadrons
$\dz$, $\dc$, $\dssp$ and $\lcp$,
and into the charmed vector meson $\dsp$
have been derived from the LEP measurements
and averaged.
The $c$ quark fragmentation fractions
represent probabilities to hadronise as
a given charmed hadron, while
the $b$ quark fragmentation fractions are defined
as sums of probabilities to produce
a particular charmed hadron
or its antiparticle.
}

\begin{document}
\maketitle
\flushbottom

\section{Introduction}
\label{sec:intro}

Charmed mesons are produced in processes of charm quark fragmentation
and in decays of bottom hadrons.
The large values of the $c$ and $b$ quark masses allow
using the perturbative QCD (pQCD) calculations
for describing the heavy quark production.
However, to describe completely the heavy quark transition into
a given hadron, a non-perturbative parametrisation tuned
to experimental results is needed.
Fragmentation functions, used
to parametrise a transfer of quark's energy to a given hadron,
are expected to be rather different for different pQCD
calculations~\cite{cacciari97,kniehl98,oleari99}.
Fragmentation fractions are probabilities for $c/b$ quark to hadronise
as a particular charmed/bottom hadron, $f(c\rightarrow H_c)$/$f(b\rightarrow H_b)$.
The fragmentation fractions of $b$ quark to a particular charmed hadron,
$f(b\rightarrow H_c)$,
can be defined similarly representing convolutions of
the $f(b\rightarrow H_b)$ fractions and branching fractions of the bottom
hadrons into a given charmed hadron.
For heavy quark ($Q$) production at large energies far from
the $Q{\bar Q}$ threshold,
the fragmentation fractions are often assumed to be universal.

Charm and bottom quark production in $e^+e^-$ annihilations
provides the cleanest environment for the heavy quark
fragmentation measurements. A compilation of charm
fragmentation fractions was performed in 1999~\cite{ctod}
using available at that time published and preliminary LEP
measurements in $Z$ decays
as well as the results obtained at centre-of-mass energies
of about $10\,$GeV.
In this article,
the fragmentation fractions of  
$c$ and $b$ quarks into the weakly decaying
charmed hadrons
$\dz$, $\dc$, $\dssp$ and $\lcp$, and into the charmed vector meson $\dsp$,
derived from the final LEP results,
are presented.
The measurements at centre-of-mass energies of $\sim10\,$GeV
are not included in this compilation
because they can be affected by exclusive production channels,
such as $e^+e^- \rightarrow D^{(*)+}D^{(*)-}$ \cite{belle_eedd},
and threshold effects.

\begin{table}
\centering
\caption{
{Charmed hadron decays and their branching fractions
used in this article.
For all decays, except $\lcpkpi$,
the current world average values~\cite{pdg2014} of the branching fractions are used.
For the $\lcpkpi$ decay, the branching fraction is obtained
as the weighted mean of the BELLE~\cite{belle_lcpkpi} and CLEO~\cite{cleo_lcpkpi} measurements.
}}
\label{tab:cdec}
\begin{tabular}{|c|c|} \hline

Charmed hadron decay & Branching fraction $~$[$\%$] \\
\hline
\hline
$\dzkpi$
& 3.88 $\pm$ 0.05~\cite{pdg2014} \\
\hline
$\dckpipi$
& 9.15 $\pm$ 0.19~\cite{pdg2014} \\
\hline
$\dssphipi$
& 2.24 $\pm$ 0.10~\cite{pdg2014} \\
\hline
$\lcpkpi$
& 6.71 $\pm$ 0.35~\cite{belle_lcpkpi,cleo_lcpkpi} \\
\hline
\hline
$\dsdz$
& 67.7 $\pm$ 0.5~\cite{pdg2014} \\
\hline
\end{tabular}
\end{table}

The LEP measurements,
used in this article,
were based on
the charmed hadron decays
summarised in table~\ref{tab:cdec}.
Results of all measurements have been updated using
new
branching fractions
of the above decays.
For all decays, except $\lcpkpi$,
the current world average values~\cite{pdg2014} of the branching fractions
are used.
For the $\lcpkpi$ decay, the branching fraction determination
in Ref.~\cite{pdg2014}
is model dependent resulting in the value
$\bran(\lcpkpi ) = (5.0\pm 1.3)\%$ with the large uncertainty
originating from the model dependence.
In this article,
the branching fraction obtained
as the weighted mean of the BELLE~\cite{belle_lcpkpi} and CLEO~\cite{cleo_lcpkpi} measurements
is used for the $\lcpkpi$ decay.
The $b$ quark fragmentation fractions are defined
as sums of probabilities to produce
a particular charmed hadron
or its antiparticle.

\section{Charm fragmentation fractions}
\label{sec:charm}

\begin{table*}
\centering
\caption{
{LEP measurements of the products of
$R_c=\Gamma(Z\rightarrow c{\bar c})/\Gamma(Z\rightarrow {\rm hadrons})$,
charm fragmentation fractions into charmed hadrons,
$f(c\rightarrow H_c)$,
and the branching fractions of the corresponding charmed hadron
decays, $\bran$.
The DELPHI and OPAL rates for $\dssp$ are corrected for the
branching fraction $\bran(\phi \rightarrow K^+K^-)$.
The first/second uncertainties are statistical/systematic.
}}
\label{tab:crates}
\begin{tabular}{|c|c|c|c|} \hline

$H_c$   & ALEPH~\cite{aleph_ctod} & DELPHI~\cite{delphi_bctod} & OPAL~\cite{opal_bctod,opal_dstar} \\
           & $\gchfb~$[$\%$] & $\gchfb~$[$\%$] & $\gchfb~$[$\%$] \\
\hline
\hline
$\dz$
& 0.370 $\pm$ 0.011 $\pm$ 0.023 & 0.3570 $\pm$ 0.0100 $\pm$ 0.0146 & 0.389 $\pm$ 0.027 $^{+0.026}_{-0.024}$ \\
\hline
$\dc$
& 0.368 $\pm$ 0.012 $\pm$ 0.020 & 0.3494 $\pm$ 0.0116 $\pm$ 0.0140 & 0.358 $\pm$ 0.046 $^{+0.025}_{-0.031}$ \\
\hline
$\dssp$
& 0.0352 $\pm$ 0.0057 $\pm$ 0.0021 & 0.0765 $\pm$ 0.0069 $\pm$ 0.0037 & 0.056 $\pm$ 0.015 $\pm$ 0.007 \\
\hline
$\lcp$
& 0.0673 $\pm$ 0.0070 $\pm$ 0.0037 & 0.0743 $\pm$ 0.0155 $\pm$ 0.0078 & 0.041 $\pm$ 0.019 $\pm$ 0.007 \\
\hline
\hline
$\dsp$
& -- & 0.1089 $\pm$ 0.0027 $\pm$ 0.0039 & 0.1041 $\pm$ 0.0020 $\pm$ 0.0040 \\
\hline

\end{tabular}
\end{table*}

\begin{table*}
\centering
\caption{
{Charm fragmentation fractions into charmed hadrons,
$f(c\rightarrow H_c)$,
derived from the LEP measurements.
The first/second uncertainties are statistical/systematic.
}}
\label{tab:cftod}
\begin{tabular}{|c|c|c|c|} \hline

$H_c$   & ALEPH~\cite{aleph_ctod} & DELPHI~\cite{delphi_bctod,delphi_dstar} & OPAL~\cite{opal_bctod,opal_dstar} \\
           & $f(c\rightarrow H_c)~$[$\%$] & $f(c\rightarrow H_c)~$[$\%$] & $f(c\rightarrow H_c)~$[$\%$] \\
\hline
\hline
$\dz$
& 55.3 $\pm$ 1.6 $\pm$ 3.4 & 53.4 $\pm$ 1.5 $\pm$ 2.2 & 58.2 $\pm$ 4.0 $^{+3.9}_{-3.6}$ \\
\hline
$\dc$
& 23.4 $\pm$ 0.8 $\pm$ 1.3 & 22.2 $\pm$ 0.7 $\pm$ 0.9 & 22.8 $\pm$ 2.9 $^{+1.6}_{-2.0}$ \\
\hline
$\dssp$
& 9.1 $\pm$ 1.5 $\pm$ 0.5 & 9.7 $\pm$ 0.9 $\pm$ 0.5 & 7.1 $\pm$ 1.9 $\pm$ 0.9 \\
\hline
$\lcp$
& 5.8 $\pm$ 0.6 $\pm$ 0.3 & 6.4 $\pm$ 1.3 $\pm$ 0.7 & 3.5 $\pm$ 1.6 $\pm$ 0.6 \\
\hline
\hline
$\dsp$, rate
& 23.3 $\pm$ 1.0 $\pm$ 0.9 & 24.1 $\pm$ 0.6 $\pm$ 0.9 & 23.0 $\pm$ 0.4 $\pm$ 0.9 \\
\hline
$\dsp$, double-tag
& -- & 25.7 $\pm$ 1.5 $\pm$ 0.6 & 22.4 $\pm$ 1.4 $\pm$ 1.4 \\
\hline

\end{tabular}
\end{table*}

Production rates of
charmed hadrons
$\dz$, $\dc$, $\dssp$ and $\lcp$
in $Z\rightarrow c{\bar c}$
were reported
by the ALEPH~\cite{aleph_ctod}, DELPHI~\cite{delphi_bctod} and OPAL~\cite{opal_bctod}
collaborations in the form of product $\gchfb$,
where $R_c=\Gamma(Z\rightarrow c{\bar c})/\Gamma(Z\rightarrow {\rm hadrons})$
and $\bran$ is the branching fraction of the corresponding charmed hadron
decay.
For $\dsp$ meson, the products were published by DELPHI~\cite{delphi_bctod}
and OPAL~\cite{opal_dstar}
(the ALEPH direct measurement of $\fcds$ is described below).
The measurements are collected in table~\ref{tab:crates}.
The DELPHI and OPAL rates for $\dssp$ were corrected for the
branching fraction $\bran(\phi \rightarrow K^+K^-)$.
The OPAL result for $\dssp$,
obtained using decay channels $\dssp \rightarrow \phi\pi^+$ and 
$\dssp \rightarrow \bar{K}^{*0} K^+$,
was expressed in terms of the decay $\dssp \rightarrow \phi\pi^+$
with the uncertainty of the ratio of the two branching fractions
included into a systematic uncertainty.

Charm fragmentation fractions are obtained
by dividing
the measured $\gchfb$ values from table~\ref{tab:crates}
by the relevant branching fractions~\cite{pdg2014} and
the Standard Model (SM) value $R_c=0.1723$~\cite{lep2013}.
The ALEPH measurement of $\dsp$ production rate
in $Z\rightarrow c{\bar c}$~\cite{aleph_ctod},
reported as
$$\fcds = 0.2333\pm0.0102 ({\rm stat})\pm0.0084 ({\rm syst}),$$
is updated using the latest values of the $\bran(D^{*+}\rightarrow D^0\pi^+)$
and $\bran(D^0\rightarrow K^-\pi^+)$ branching fractions~\cite{pdg2014}.

The double-tag measurements of the product
$f(c\rightarrow D^{*+}) \cdot \bran(\dsp \rightarrow\dz\pi^+)$
were performed by DELPHI~\cite{delphi_dstar} and OPAL~\cite{opal_dstar}.
The double-tag method, based on
tagging of one heavy quark
and
detection of 
a low momentum
charged pion from the $\dsp \rightarrow \dz \pi^+$ decay
in the hemisphere opposite to the tagged quark,
substantially reduces backgrounds, although at the price of reduced
statistics.
The $\fcds$ values derived from the product by DELPHI,
$$\fcds = 0.255\pm0.015 ({\rm stat})\pm0.006 ({\rm syst}),$$
and by OPAL,
$$\fcds = 0.222\pm0.014 ({\rm stat})\pm0.014 ({\rm syst}),$$
are updated using the latest value of the $\bran(D^{*+}\rightarrow D^0\pi^+)$
branching fraction~\cite{pdg2014}.

Charm fragmentation fractions derived from the LEP measurements
are collected in table~\ref{tab:cftod}.

\section{Bottom fragmentation fractions}
\label{sec:bottom}

\begin{table*}
\centering
\caption{
{LEP measurements of the
bottom fragmentation fractions into charmed hadrons,
$f(b\rightarrow H_c)$, and products of
$R_b=\Gamma(Z\rightarrow b{\bar b})/\Gamma(Z\rightarrow {\rm hadrons})$,
$f(b\rightarrow H_c)$
and the branching fractions of the corresponding charmed hadron
decays, $\bran$.
The DELPHI and OPAL rates for $\dssp$ are corrected for the
branching fraction $\bran(\phi \rightarrow K^+K^-)$.
The first/second uncertainties are statistical/systematic.
}}
\label{tab:brates}
\begin{tabular}{|c|c|c|c|} \hline

$H_c$   & ALEPH~\cite{aleph_btod} & DELPHI~\cite{delphi_bctod} & OPAL~\cite{opal_bctod,opal_dstar} \\
           & $f(b\rightarrow H_c)~$[$\%$] & $\gbhfb~$[$\%$] & $\gbhfb~$[$\%$] \\
\hline
\hline
$\dz$
& 60.5 $\pm$ 2.4 $\pm$ 1.6 & 0.4992 $\pm$ 0.0162 $\pm$ 0.0304 & 0.454 $\pm$ 0.023 $^{+0.025}_{-0.026}$ \\
\hline
$\dc$
& 23.4 $\pm$ 1.3 $\pm$ 1.0 & 0.4525 $\pm$ 0.0204 $\pm$ 0.0226 & 0.379 $\pm$ 0.031 $^{+0.028}_{-0.025}$ \\
\hline
$\dssp$
& 18.3 $\pm$ 1.9 $\pm$ 0.9 & 0.1259 $\pm$ 0.0100 $\pm$ 0.0063 & 0.166 $\pm$ 0.018 $\pm$ 0.016 \\
\hline
$\lcp$
& 11.0 $\pm$ 1.4 $\pm$ 0.6 & 0.0962 $\pm$ 0.0187 $\pm$ 0.0083 & 0.122 $\pm$ 0.023 $\pm$ 0.010 \\
\hline
\hline
$\dsp$
& -- & 0.1315 $\pm$ 0.0035 $\pm$ 0.0053 & 0.1334 $\pm$ 0.0049 $\pm$ 0.0078 \\
\hline

\end{tabular}
\end{table*}

\begin{table*}
\centering
\caption{
{Bottom fragmentation fractions into charmed hadrons,
$f(b\rightarrow H_c)$,
derived from the LEP measurements.
The first/second uncertainties are statistical/systematic.
}}
\label{tab:bftod}
\begin{tabular}{|c|c|c|c|} \hline

$H_c$   & ALEPH~\cite{aleph_btod} & DELPHI~\cite{delphi_bctod,delphi_dstar} & OPAL~\cite{opal_bctod,opal_dstar} \\
           & $f(b\rightarrow H_c)~$[$\%$] & $f(b\rightarrow H_c)~$[$\%$] & $f(b\rightarrow H_c)~$[$\%$] \\
\hline
\hline
$\dz$
& 59.7 $\pm$ 2.4 $\pm$ 1.3 & 59.6 $\pm$ 1.9 $\pm$ 3.6 & 54.2 $\pm$ 2.7 $^{+3.0}_{-3.1}$ \\
\hline
$\dc$
& 23.3 $\pm$ 1.3 $\pm$ 1.0 & 23.0 $\pm$ 1.0 $\pm$ 1.1 & 19.2 $\pm$ 1.6 $^{+1.4}_{-1.3}$ \\
\hline
$\dssp$
& 14.4 $\pm$ 1.5 $\pm$ 0.8 & 12.8 $\pm$ 1.0 $\pm$ 0.6 & 16.6 $\pm$ 1.8 $\pm$ 1.6 \\
\hline
$\lcp$
& 7.2 $\pm$ 0.9 $\pm$ 0.6 & 6.6 $\pm$ 1.3 $\pm$ 0.6 & 11.3 $\pm$ 2.1 $\pm$ 0.9 \\
\hline
\hline
$\dsp$, rate
& -- & 23.2 $\pm$ 0.6 $\pm$ 0.9 & 23.5 $\pm$ 0.9 $\pm$ 1.4 \\
\hline
$\dsp$, double-tag
& -- & -- & 17.5 $\pm$ 1.6 $\pm$ 1.2 \\
\hline

\end{tabular}
\end{table*}

Production rates of
charmed hadrons
$\dz$, $\dc$, $\dssp$ and $\lcp$
in $Z\rightarrow b{\bar b}$
were measured
by ALEPH~\cite{aleph_btod}, DELPHI~\cite{delphi_bctod} and OPAL~\cite{opal_bctod}.
The ALEPH collaboration converted the measured rates
into the bottom fragmentation
fractions, while the DELPHI and OPAL measurements were reported
in the form of product $\gbhfb$,
where $R_b=\Gamma(Z\rightarrow b{\bar b})/\Gamma(Z\rightarrow {\rm hadrons})$
and $\bran$ is the branching fraction of the corresponding charmed hadron
decay.
For $\dsp$ meson, the products were published by DELPHI~\cite{delphi_bctod}
and OPAL~\cite{opal_dstar}.
The measurements are collected in table~\ref{tab:brates}.
The DELPHI and OPAL rates for $\dssp$ were corrected for the
branching fraction $\bran(\phi \rightarrow K^+K^-)$.
The OPAL result for $\dssp$,
obtained using decay channels $\dssp \rightarrow \phi\pi^+$ and 
$\dssp \rightarrow \bar{K}^{*0} K^+$,
was expressed in terms of the decay $\dssp \rightarrow \phi\pi^+$
with the uncertainty of the ratio of the two branching fractions
included into a systematic uncertainty.

The $f(b\rightarrow H_c)$ values measured by ALEPH
are updated using the latest values
of the relevant branching fractions~\cite{pdg2014}.
For DELPHI and OPAL,
the bottom fragmentation fractions are obtained
by dividing
the measured $\gbhfb$ values from table~\ref{tab:brates}
by the relevant branching fractions~\cite{pdg2014} and
the SM value $R_b=0.21579$~\cite{lep2013}.

The double-tag measurement of the product
$f(b\rightarrow D^{*+}) \cdot \bran(\dsp \rightarrow\dz\pi^+)$
was performed by OPAL~\cite{opal_dstar}.
The $\fbds$ value derived from the product,
$$\fbds = 0.173\pm0.016 ({\rm stat})\pm0.012 ({\rm syst}),$$
is updated using the latest value of the $\bran(D^{*+}\rightarrow D^0\pi^+)$
branching fraction~\cite{pdg2014}.

Bottom fragmentation fractions derived from the LEP measurements
are collected in table~\ref{tab:bftod}.
In ref.~\cite{aleph_ctod}, ALEPH reported the ratio
$$\frac{R_b \cdot f(b\rightarrow D^{*+})}{R_c \cdot f(c\rightarrow D^{*+})} = 1.15\pm 0.06.$$
Using the ratio, the $f(c\rightarrow D^{*+})$ value derived from the ALEPH $\dsp$ rate measurement, and the SM values of $R_c$ and $R_b$ gives $f(b\rightarrow D^{*+})=21.4\%$
in general agreement with other results.
However,
this estimate is not included in the calculation
of the LEP average values
because there is not enough information in ref.~\cite{aleph_ctod}
for evaluation of its uncertainties.

\section{Average LEP fragmentation fractions}
\label{sec:average}

For each fragmentation fraction considered,
the results are averaged 
using a standard weighted least-squares procedure~\cite{pdg2014}.
The statistical and systematic uncertainties
are added in quadrature and the combined
uncertainties are used.
For $\dz$ and $\dc$ from OPAL, the systematic uncertainties are
slightly asymmetric and
the uncertainty with a larger absolute value is used.
The systematic uncertainties related to the detector description,
signal-extraction procedure and Monte Carlo statistics are assumed
to be uncorrelated between the LEP experiments.
The uncertainties related to the modelling of $Z\rightarrow c{\bar c}$
and  $Z\rightarrow b{\bar b}$ events are taken as the correlated uncertainties.
Taking the correlations into account produces small but visible effects
on the resulting values and their combined uncertainties.

\begin{table}
\centering
\caption{
{Average values of the fragmentation fractions into charmed hadrons,
derived from the LEP measurements,
for charm quark,
$f(c\rightarrow H_c)$, and
bottom quark,
$f(b\rightarrow H_c)$.
The first uncertainties are the combined statistical and systematic
uncertainties of the LEP measurements.
The second uncertainties originate from the limited knowledge
of the relevant branching fractions~\cite{pdg2014,belle_lcpkpi,cleo_lcpkpi}.
}}
\label{tab:cblep}
\begin{tabular}{|c|c|c|} \hline

$H_c$   & $f(c\rightarrow H_c)~$[$\%$] & $f(b\rightarrow H_c)~$[$\%$] \\
\hline
\hline
$\dz$
& 54.2 $\pm$ 2.4 $\pm$ 0.7 & 58.7 $\pm$ 2.1 $\pm$ 0.8 \\
\hline
$\dc$
& 22.5 $\pm$ 1.0 $\pm$ 0.5 & 22.3 $\pm$ 1.1 $\pm$ 0.5 \\
\hline
$\dssp$
& 9.2 $\pm$ 0.8 $\pm$ 0.5 & 13.8 $\pm$ 0.9 $\pm$ 0.6 \\
\hline
$\lcp$
& 5.7 $\pm$ 0.6 $\pm$ 0.3 & 7.3 $\pm$ 0.8 $\pm$ 0.4 \\
\hline
\hline
$\dsp$, rate
& 23.4 $\pm$ 0.7 $\pm$ 0.3 & 23.3 $\pm$ 1.0 $\pm$ 0.3 \\
\hline
$\dsp$, double-tag
& 24.4 $\pm$ 1.3 $\pm$ 0.2 & 17.5 $\pm$ 2.0 $\pm$ 0.1 \\
\hline
\hline
$\dsp$, combined
& 23.6 $\pm$ 0.6 $\pm$ 0.3 & 22.1 $\pm$ 0.9 $\pm$ 0.3 \\
\hline
\end{tabular}
\end{table}

Average values of the charm and bottom fragmentation fractions
into charmed hadrons,
derived from the LEP measurements,
are collected in table~\ref{tab:cblep}.
The sum of the average $f(c\rightarrow H_c)$ fragmentation
fractions
into
the weakly decaying
charmed hadrons
$\dz$, $\dc$, $\dssp$ and $\lcp$
is
$$91.6\pm3.3 ({\rm stat\oplus syst})\pm1.0 ({\rm branching ~fractions})\%$$
leaving a large room for the fragmentation
of a charm quark into $\Xi_c^+$, $\Xi_c^0$ and $\Omega_c^0$ baryons.
The sum of the average $f(b\rightarrow H_c)$ fragmentation
fractions
into
the weakly decaying
charmed hadrons
$\dz$, $\dc$, $\dssp$ and $\lcp$
is
$$102.1\pm3.1 ({\rm stat\oplus syst})\pm1.1 ({\rm branching ~fractions})\%$$
reflecting the fact that more than
one charm quark is produced on average in the hadronisation
and decay
of one bottom quark
due to $b\rightarrow c{\bar c}s$ processes.
The sums uncertainties
are calculated taking into account correlations between systematic
uncertainties of the four fragmentation fractions in each
measurement~\cite{aleph_ctod,delphi_bctod,opal_bctod,aleph_btod}.

\section{Summary}
\label{sec:summary}

The fragmentation fractions of  
$c$ and $b$ quarks into
the weakly decaying
charmed hadrons
$\dz$, $\dc$, $\dssp$ and $\lcp$,
and into the charmed vector meson $\dsp$
have been derived from the LEP measurements
and averaged.
The $b$ quark fragmentation fractions are defined
as sums of probabilities to produce
a particular charmed hadron
or its antiparticle.
The average values obtained
are intended for normalisation of the non-perturbative
component of charm and bottom hadronisation
in analytic and Monte Carlo calculations.

\end{document}